\def\approxlt{\lower.2em\hbox{$\buildrel < \over \sim$}}
\def\approxgt{\lower.2em\hbox{$\buildrel > \over \sim$}}
\def \kms{\hbox{km$\,$s$^{-1}$}}
\def \ls{\hbox{L$_{\odot}$}}
\def \ms{\hbox{M$_{\odot}$}}
\def \etal{{et~al. }}
\def \eg{{e.g.}}

\def\lir{{\hbox {$L_{\rm IR}$}}}
\def\lco{{\hbox {$L_{\rm CO}$}}}
\documentstyle[12pt,aasms4]{article}
\begin{document}

\title{Molecular Gas Depletion and Starbursts \\ in Luminous Infrared
Galaxy Mergers}

\author{Yu Gao$^{1,2,3}$  \ and \ 
Philip M. Solomon$^1$}
\affil{1 Dept. of Physics \& Astronomy, State University of New York, 
Stony Brook, NY 11794 \\ 2 Laboratory for Astronomical Imaging,
University of Illinois, Urbana, IL 61801 \\ 
3 current address: Dept. of Astronomy, 
University of Toronto, Toronto, ON M5S 3H8 CANADA
}
\authoremail{gao@astro.utoronto.ca, psolomon@astro.sunysb.edu}
 
\begin{abstract}

Most luminous infrared galaxies (LIGs) are closely interacting/merging
systems rich in molecular gas.  We study
here the relationship between  the  stage of the galaxy-galaxy 
interactions, the molecular gas mass, and the  star
formation rate as deduced  from the infrared luminosity $L_{\rm IR}$ in LIGs.  
We find a correlation between the CO(1-0)
luminosity  (a measure of molecular mass M(H$_2$))  and the projected
separation of merger nuclei (indicator of
merging stages)  in a sample of 50 LIG mergers,  which shows that 
the molecular gas content decreases as
merging advances.  The starburst is due to enhanced star formation 
in pre--existing molecular
clouds  and not to the formation of more molecular clouds from HI.   
The molecular content is being  rapidly depleted due to the
 starbursts as merging progresses.   This is further supported by 
an anti-correlation between
$L_{\rm IR}/M$(H$_2$), the global measure of the star formation rate 
per unit gas mass, and the projected
separation  implying  an enhanced  star formation ``efficiency''  
in late stage mergers compared to that of early
mergers.  This is the first evidence connecting the depletion of 
molecular gas with starbursts in interacting galaxies. 

\end{abstract}


\section{INTRODUCTION}

Luminous infrared galaxies (LIGs, $L_{\rm IR} \approxgt 2\times10^{11}~\ls$, 
$H_0=75$~\kms~Mpc$^{-1}$), are the dominant population among objects with
$L_{\rm bol} \approxgt 2\times 10^{11} \ \ls$ in the local
universe (see the comprehensive review by Sanders \& Mirabel 1996, also
for a definition of $L_{\rm IR}$). 
Most LIGs are merging/interacting (\eg, Sanders 1992; 
Leech \etal 1994; Murphy \etal 1996) and molecular 
gas rich with a high star formation (SF) rate as most of the 
luminosity is radiated by dust in the far-IR.
Sanders \etal (1988) suggested that the most extreme
ultra-LIGs (ULIGs) with 
$L_{\rm IR} \approxgt 10^{12} \ \ls$ comparable with
the bolometric luminosity of quasars (QSOs), are dust-enshrouded QSOs
triggered by the interaction/merger of two {\it gas-rich} spirals.
However, a starburst origin 
(\eg, Joseph \& Wright 1985) even for ULIGs
has also been suggested. Evidence in favor of a starburst includes 
the H$_2$ gas  content, radio 
continuum morphology and ISO mid-IR
spectroscopy (\eg, Solomon \etal 1997 (S97); Downes \& Solomon 1998 (DS98);
Condon \etal 1991; Genzel \etal 1998). 
The observed abundant supply of {\it dense} H$_2$ gas traced by HCN 
emission, usually found only in
SF cores (Solomon \etal 1992; Gao 1996; Gao \& Solomon 1999) shows
that ULIGs are ideal stellar nurseries.

Although there have been extensive CO observations in LIGs/ULIGs
including CO imaging (\eg, Scoville \etal 1991; 1997; DS98 ),
there  has been no systematic attempt to trace the H$_2$ gas 
properties during the galaxy-galaxy merger sequence starting with 
systems roughly separated by a galactic diameter. 
The purpose of this study is to investigate the connection between 
the progress of a merger and the H$_2$ gas responsible for 
the SF. There are many close mergers 
where the progenitors are not gas-rich galaxies; 
these may never become LIGs. All LIGs/ULIGs are H$_2$ gas rich 
(Sanders \etal 1991; Gao 1996; S97). 

The ideal sample to trace a merger sequence should contain galaxies of 
various merging stages that initially started 
with roughly comparable molecular gas content. Since this is impossible, 
we select LIGs where 
we can identify the 2 progenitors from CCD images usually in 
the red or near-IR. By measuring the CO(1-0) luminosity as a function of the
merger separation, the time dependence of the gas 
content can be traced statistically.

\section{THE SAMPLE}

We focus exclusively on mergers 
satisfying $\ S_{\rm sep} \approxlt (D_1+D_2)/2$,\  where $S_{\rm sep}$ 
is the projected separation between the nuclei of the merging galaxies of
the major diameters $D_1$ and $D_2$.
In reality, we require the pairs have 
touching/overlapping optical disks/tails.
We exclude late mergers of $S_{\rm sep}\approxlt 2 ''$ ($\sim 1$ kpc 
at an angular distance of $\sim 100$ Mpc) so that $S_{\rm sep}$ 
can be reliably measured.  This also helps to 
distinguish genuine mergers from projection effects, 
exclude galaxies where there is confusion of the extranuclear 
starburst regions with galactic nuclei, and minimize
other potential biases introduced by the seeing limit or severe dust 
obscuration. At larger distance, LIGs with $S_{\rm sep}\sim1$~kpc
similar to the nearest ULIGs Arp~220, Mrk~273 
will be simply indistinguishable from
the single-nucleus galaxies without sub-arcsec imaging. Even at moderate 
distances ($\sim 200$ Mpc), some previously accepted close doubles,
\eg, Mrk~231 (Armus \etal 1994) and IRAS~12112+0305 (Sanders 1992) turn out
to be singles as seen by HST (Surace \etal 1998; Sanders 1998, private 
communications).

Our sample selection is simple: all LIGs with \ 
$L_{\rm IR}~\approxgt~ 2\times 10^{11}~\ls, \ 
2'' \approxlt \ S_{\rm sep} \approxlt (D_1+D_2)/2$ \ 
are selected if they have both CO data and CCD images. 
A heterogeneous sample of 50 LIGs has been used (Table~1);
19 of them are ULIGs. 


CO(1-0) observations have been conducted in $\sim 30$ LIGs of mostly
mergers (Gao 1996). Most have been observed with the NRAO 12m 
and a representation is shown in Fig.~1 (details
will be discussed in a future paper).
This nearly completes the CO data for nearby LIG 
mergers in the Bright Galaxy Sample (BGS, Soifer \etal 1989;
Sanders \etal 1995). 

A {\it volume-limited} ($cz \approxlt 12,000 \kms$) nearly complete
sample of 20 LIGs, mainly from the BGS, is highlighted in Table~1. 
Most LIGs but not ULIGs are in this sub-sample.


\section{$L_{\rm CO}$---$S_{\rm sep}$ CORRELATION}

We have separated the sample into 31 LIGs with
$2\times 10^{11} < L_{\rm IR} \approxlt 10^{12}~\ls$ and 19 ULIGs 
with $L_{\rm IR}>10^{12}$\ls.
The observed CO luminosities range over an order of magnitude
$log (\lco/L_l) =$9.2--10.3, $L_l$=K \kms pc$^2$ for the entire sample.
This is at the high end of the distribution for
normal molecular gas-rich galaxies. 
However for ULIGs the range is smaller with 
$<log(\lco/L_l)> = 9.92 \pm 0.12$ (S97). 

The correlation between CO luminosity $L_{\rm CO}$ and the nuclei separation $S_{\rm sep}$
is evident in {\bf Fig.~2a}. We compute correlation coefficients of 
$\gamma=0.80$ for 31 LIGs, and $\gamma=0.78$ for 20 {\it volume-limited}
LIGs, with fits:
\begin{equation}
L_{\rm CO}=10^{9.0} S^{0.80\pm 0.18}_{\rm sep}, \ \ \
L_{\rm CO}=10^{9.1} S^{0.82\pm 0.20}_{\rm sep}
\end{equation}
respectively. Clearly, this correlation indicates
$L_{\rm CO}$ is decreasing as merging progresses to advanced stages. 
The system with the highest $L_{\rm CO}$ in the sample is Arp~302 which  
also has the largest separation. Even if Arp~302 is excluded 
the correlation remains the same. While the correlation for the entire
heterogeneous sample is still obvious ($\gamma=0.6$) the correlation 
for ULIGs alone (open circles in Fig.~2) is not significant ($\gamma=0.2$).
In view of the near constancy of 
\lco \ for ULIGs (S97) this is not surprising.


There is no correlation between IR luminosity \lir \ and 
separation $S_{\rm sep}$ 
(Fig.~2b) indicating the IR emission is not enhanced as the merger
progresses, at least prior to reaching 
$S_{\rm sep}\approxlt 2$ kpc. As discussed above we have
excluded close mergers/singles. Many of these are ULIGs. An anti-correlation 
of $L_{\rm IR}$ with $S_{\rm sep}$ appears to hold
if small separation ULIGs
are included, particularly 10 ULIGs in the BGS (Sanders \etal 1988).

There is a significant anti-correlation between $L_{\rm IR}/M({\rm H_2})$
(or $L_{\rm IR}/L_{\rm CO}$, a SF ``efficiency'' (SFE)
measure of the global
SF rate per solar mass of gas) and $S_{\rm sep}$ 
(Fig.~2c, $\gamma=-0.61$ for 31 LIGs). Here we have used a standard
conversion factor of $4.7 \ms/L_l$ applicable to giant molecular clouds (GMCs)
to obtain H$_2$ mass from \lco.
It clearly suggests that the SF rate, normalized to the 
available molecular mass, or SFE, increases at
small separations. This indicates an increased rate 
for the conversion of gas to stars, which is
the primary criterion for a starburst.

We checked for  possible selection effects
by examining the dependence of the parameters on distance to the source.
There is no strong dependence of the projected
separation upon the distance even in the large
heterogeneous sample, the correlation 
between \lco \ and the distance is weak and 
is nonexistent for the nearby BGS LIGs. 

\section{IMPLICATIONS AND DISCUSSION}

We conclude that there is a strong correlation between \lco \ and 
$S_{\rm sep}$ for LIGs although almost no correlation for ULIGs. 
CO observations of widely separated interacting
galaxies ($\ S_{\rm sep} \approxgt (D_1+D_2)/2$, {\bf non-mergers})
show no correlation between \lco \ and $S_{\rm sep}$ 
(Combes \etal 1994). 
After merging begins $\ S_{\rm sep} \approxlt (D_1+D_2)/2 \sim 20$ kpc,
the molecular gas content is decreasing as
merging advances. Since we may overestimate the H$_2$ mass in ULIGs and
late mergers (S97; DS98) by using a constant conversion factor,
the dependence of H$_2$ mass on separation may be even steeper than 
that in Eq.~1.

Mergers separated by a few kpc produce the {\bf same \lir } as 
mergers with $\ S_{\rm sep} \sim15$~kpc {\bf with less H$_2$ gas}
(Fig.~2 \& Eq.~1). It is not the increase in molecular gas that
leads to the high \lir \ but the increase in average gas surface
density which produces a higher SFE. 
In more than a dozen early/intermediate stage
mergers in this sample, CO imaging reveals that most of the 
gas is in the inner merging disks 
except for widely separated pairs not yet closely
interacting,  which have extended CO (Gao \etal 1997; 1998a;
Lo, Gao \& Gruendl 1997). Arp~302 (Lo \etal 1997), 
the widest pair ($S_{\rm sep}$=25~kpc), has two extended
CO disks ($>10$ kpc) and one of them shows roughly an exponential 
disk with a scale length of $\sim 7$ kpc
extending out to 20 kpc in diameter. NGC~6670 ($S_{\rm sep}=15$ kpc)
shows central sources with $\approxgt 1$ kpc radii and strong CO disks
out to $\sim 5$ kpc with weaker emission out to 15 kpc in both galaxies.
Arp~238 ($S_{\rm sep}$=12~kpc) and Arp~55 ($S_{\rm sep}$=10~kpc), the
intermediate stage mergers, show central $\sim 1$--2 kpc sources and
extensions out to $\sim 5$--10 kpc in both pairs of disks. Late mergers
like Arp~299 (Aalto \etal 1997) and NGC~6090 ($S_{\rm sep}<5$ kpc) show 
gas concentrations between the merging disks. 

For a pair of gas-rich spirals initially separated at 
$S_{\rm sep}^0 \approxlt 20$ kpc, 
merging at (orbital decay speed) $V\sim 40 \kms$, the time to reach 
$S_{\rm sep}^{\rm min}\approxgt 2 {\rm kpc}$ is \ 
$500 \times ({{S_{\rm sep}^0-S_{\rm sep}^{\rm min}}\over{20 \rm kpc}}) 
({{40 \kms}\over V})$ Myr.
If the LIGs with different separations are interpreted 
as a sequence then, using a standard conversion factor and Eq.~1,
the implied average SF (gas depletion) rate is
\begin{equation}
100 \times A \times ({V\over{40 \kms}})
({{20\rm kpc}\over {S_{\rm sep}^0}})^{0.2} \ \ms/yr
\end{equation}
where 
$A=(1-({{S_{\rm sep}^{\rm min}}\over{S_{\rm sep}^0}})^{0.8})/(1-{{S_{\rm sep}^{\rm min}}\over{S_{\rm sep}^0}})\approxlt 1$. 
Of course, the trend evident in Fig.~2a may not be a true time sequence 
since it is a snapshot of many LIGs at various
separations. Nevertheless, this SF rate
deduced from the observed  depletion of H$_2$ gas is similar to that 
inferred from $L_{\rm IR}$ (under the assumption that 
the IR luminosity is from imbedded high mass stars):
about 100 \ms/yr for 
$L_{\rm IR} \sim 5 \times 10^{11} \ls$ (\eg, Sanders \& Mirabel 1996).

At the extreme end of the merger sequence are Ultra-LIGs (ULIGs) 
with small  separations or no evidence
of 2 remaining nuclei, which have very compact gas concentrations 
and about 5 times more \lir \ than LIGs.  Most of the
molecular gas in ULIGs is concentrated in the central 1 kpc 
and is not in normal GMCs; the translation of \lco \ to H$_2$ mass is
fundamentally different (S97; DS98) and the H$_2$ mass is lower than 
expected from use of a standard coversion factor. 
Nevertheless, ULIGs still have close to $10^{10} \ms$ H$_2$.  Even
allowing for this, ULIGs do not follow the correlations we find 
for LIGs (Fig.~2). The difference 
may be due to the different physical
parameters of the starbursts. The starbursts in ULIGs appear to take 
place predominantly in {\bf extreme starburst} regions
of $\sim 200$ pc size and $\lir \sim 3\times 10^{11} \ls$ (DS98) with 
substantial velocity structure characterized by rotation or bar-like
motion. They are the primary sources of the millimeter continuum 
with more than 1,000 magnitudes of visual extinction. For example,
Arp 220 has two extreme starburst regions. These may be the
remnants of nuclei or condensations in the gas but they now are composed 
almost entirely of gas and young stars. AGNs might also contribute to
\lir \ in some ULIGs (\eg, Veilleux \etal 1997). 

Mihos \& Hernquist (1996) have predicted the SF 
rate and the gas depletion during a merger sequence in 
two extreme cases: galaxies with bulges and without (see their Fig.~5).
The  decrease in H$_2$ mass in our sample (Fig.~2a) 
is consistent with their model prediction (especially when ULIGs
are excluded) for a diverse group of initial conditions and bulge/disk ratios.
 In the models only mergers with bulges 
have enough gas left for a final extreme burst as in ULIGs.

The models failed to predict gas concentration in the overlap
region, either HI in early mergers, \eg, NGC~6670 (Wang \etal 1998) or 
H$_2$ in intermediate mergers, \eg, Arp~244 and Arp~299 (Gao \etal 1998b;
Aalto \etal 1997). Although gas inflow
into the center in late mergers is predicted (Mihos \& 
Hernquist 1996; Barnes \& Hernquist 1996), this is near the final merging 
($S_{\rm sep} \approxlt $ 3 kpc). 
Successful models should distinguish between diffuse HI and 
H$_2$ in GMCs and produce this feature
when galaxies are still 5--15 kpc apart. 
SF in the overlap region is predicted in a model by the compression
of pre-existing GMCs due to the heating of the atomic ISM by shocks
(Jog \& Solomon 1992). Pre-existing GMCs are clearly the source
of SF even before the true merging of the disks as is evident
in Arp~302.

\section{CONCLUDING REMARKS}

We show here that there is 
a strong correlation between CO luminosity, a measure of M(H$_2$), and
the projected separation for luminous infrared galaxies (LIGs). 
Although there is no correlation between the star formation (SF) 
rate (\lir) and separation, LIGs with smaller separation have
a higher SF rate per solar mass of H$_2$. This increase 
in SF ``efficiency'' is due to increased H$_2$ gas surface density.
Advanced mergers have {\bf less} H$_2$ mass (not more) than when the
mergers started. This implies that the starburst is not fueled by a large
inflow of a huge mass of diffuse HI gas from large galactic radii converting
to H$_2$, but rather that existing GMCs form stars throughout
the merging process over a few hundred million years. 

Although the correlations are robust for LIGs they do not exist for 
ultra-LIGs (ULIGs). The difference may be due to the 
different physical parameters and timescales of the starbursts. 
The starbursts in ULIGs 
appear to take  place predominantly  in  {\bf extreme starburst} regions
with radii $\sim 100$ pc, $\sim 10^9$ \ms H$_2$, 
$\lir \sim 3\times 10^{11} \ls$ and a very high SF efficiency (DS98).
The dynamical timescales in extreme starbursts
are characterized by only about a million years ($V\sim 250 \kms$ and 
size $\sim 200$ pc).

\acknowledgements

We thank D.C. Kim, D.B. Sanders and J.D. Goldader for the use of some optical
images. Helpful suggestions and discussions with D.B. Sanders 
are gratefully acknowledged. We appreciate very much for the thorough review
and thoughtful/helpful comments of the referee, T.W. Murphy, Jr. 
YG's research at LAI is funded by NSF grant AST96-13999 and U. of Illinois.

\clearpage

\figcaption{A representation of the CO observations made with the 
NRAO 12m (FWHM beam 55$''$ is indicated). The vertical scale of the CO
spectra is the antenna temperature T$_R^*$ in mK, the horizontal axis is
the redshift in \kms. The total CO luminosity in NGC~6670 is $\sim 70\%$
of the sum of the two beam measurements.}

\figcaption{(a) (top) CO luminosity vs. $S_{\rm sep}$ for 50 LIG mergers.
Strong correlation holds 
for less luminous LIGs ($L_{\rm IR}\approxlt 10^{12} \ls$, filled circles
with line fit),
which resemble the {\it volume-limited} complete sample, whereas 
no correlation for ultraluminous ones (ULIGs, open circles). 
(b) (middle) No correlation between $L_{\rm IR}$ and $S_{\rm sep}$ for
either ULIGs or LIGs.
(c) (bottom) Anti-correlation between $L_{\rm IR}/M$(H$_2$), the measure
of star formation efficiency (SFE), 
and $S_{\rm sep}$ suggests the enhanced SFE of late-stage mergers. 
We use the same conversion factor from \lco \ to H$_2$ mass
even though there is evidence that ULIGs require a smaller conversion 
factor (S97; DS98). The fit is for LIGs.}

\clearpage
 
\scriptsize
\begin{deluxetable}{lrcrr}
\tablenum{1}
\tablewidth{0pt}
\tablecaption{Properties of Luminous Infrared Galaxy Mergers}
\tablehead{
\colhead{Source/Name}           &
\colhead{cz}             & \colhead{$L_{\rm IR}$}   &
\colhead{$L_{\rm CO}$\tablenotemark{a}}             & 
\colhead{Separation\tablenotemark{b}}               \\
\colhead{IRAS}                      &
\colhead{\kms}           & \colhead{$10^{11}~\ls$}  & 
\colhead{$10^9~L_l$}                          & 
\colhead{$''$, kpc}} 

\startdata

$00057+4021$        & 13516 &  4.3 & 3.8$^B$ & 2.5,  2.1$^1$\nl
$00188-0856$        & 38550 & 27.8 & 6.8$^B$ & 6.8, 13.8$^{1,5}$\nl
$00267+3016$/Mrk551  & 15119 &  6.3 & 8.4$^A$ & 11.0,  9.9$^2$\nl
{\bf Arp236,VV114,IC1623} &  {\bf 6016} & {\bf 4.7} & {\bf 10.7$^A$} & 
{\bf 16.0, 6.1}$^3$ \nl
{\bf 01077$-$1707} & {\bf 10540} & {\bf 4.2} & {\bf 8.4$^A$} & 
{\bf 35, 22.5}$^3$  \nl
{\bf 01418$+$1651/IIIZw35}     &  {\bf 8215} & {\bf 4.2} & {\bf 2.1$^C$} & 
{\bf 8.2, 4.1}$^1$  \nl
{\bf 02114$+$0456/Mrk1027}      &  {\bf 8913} & {\bf 2.5} & {\bf 5.4$^A$} & 
{\bf 9.0, 5.0}$^2$  \nl
$02483+4302$ & 15571 & 6.2 & 2.9$^B$ & 3.8, 3.6$^4$ \nl
{\bf 02512$+$1446/UGC2369}     &  {\bf 9354} & {\bf 3.9} & {\bf 7.2$^A$} & 
{\bf 22.5, 13.1}$^{3,9}$\nl
{\bf 03359$+$1523}             & {\bf 10600} & {\bf 3.3} & {\bf 6.9$^C$} & 
{\bf 10.0, 6.5}$^3$ \nl
$03521+0028$        & 45530 & 30.2 & 9.3$^B$ & 1.6,  3.6$^5$ \nl
$04232+1436$        & 23855 & 11.1 & 9.0$^B$ & 4.6,  6.3$^9$ \nl
$04473+8529$        & 38383 &  4.2 & 8.0$^A$ & 5.0, 10.1$^6$ \nl
$06035-7102$        & 23823 & 15.2 & 8.4$^D$ & \nodata, 9.0$^7$\nl
$06206-6315$        & 27713 & 16.5 &11.2$^D$ & \nodata, 5.0$^7$\nl
$08572+3915$        & 17480 & 11.9 & 1.7$^A$ & 5.5, 4.6$^{3,8}$ \nl
{\bf Arp55,UGC4881}   & {\bf 11773} & {\bf 4.7} & {\bf 12.0$^C$} & 
{\bf 15.0, 10.7}$^3$ \nl
$10035+4852$ & 19427 & 9.3 & 7.0$^B$ &  8.7, 9.8$^{9}$ \nl
{\bf 10039$-$3338/IC2545}       & {\bf 10235} & {\bf 4.5} & {\bf 2.9$^E$} &
{\bf 5.0, 3.1}$^{E,10}$ \nl
$10190+1322$        & 22987 & 11.3 & 8.2$^B$ & 5.0, 6.6$^{9}$ \nl
$10565+2448$        & 12926  & 9.6 & 5.8$^B$ & 8.0, 6.2$^{3,5}$\nl
{\bf Arp299,NGC3690}  & {\bf  3115} & {\bf 6.4} & {\bf 3.0$^C$} & 
{\bf 25.0, 4.5}$^{2,3}$  \nl
$11506+1331$        & 38206 & 25.9 & 7.5$^B$ & 4.5, 9.0$^9$ \nl
$12593+6516$/Mrk238 & 14949 & 2.5 & 6.8$^A$ & 14.8, 13.2$^3$\nl
{\bf 13001$-$2339}    & {\bf  6446} & {\bf 2.4} & {\bf 2.6$^D$} & 
{\bf 6.5, 2.6$^{10}$} \nl
$13106-0922$        & 52330 & 17.8 & 8.9$^B$ & 2.0, 5.6$^9$ \nl
{\bf Arp238,UGC8335}  & {\bf  9453} & {\bf 5.2} & {\bf 4.2$^C$} & 
{\bf 21.0, 12.1}$^{1,3}$ \nl
{\bf NGC5256,Mrk266} & {\bf  8239} & {\bf 3.1} & {\bf 5.7$^C$} & 
{\bf 9.5,  4.8}$^2$      \nl
$13452+1232$/4C12.5 & 36520 & 21.1 & 11.2$^F$& 2.0, 3.9$^{5,8}$ \nl
$13536+1836$/Mrk463  & 15140 &  5.3 & 2.8$^G$ &  3.9, 3.5$^{2,8}$ \nl
{\bf Mrk673,IC4395}  & {\bf 10980} & {\bf 2.2} & {\bf 4.0$^A$} &
{\bf 6.0, 4.0}$^2$ \nl
$14348-1447$        & 24677 & 20.4 & 14.0$^C$&  4.5, 6.5$^3$    \nl
{\bf Arp302,UGC9618}  & {\bf 10166} & {\bf 4.1} & {\bf 17.0$^A$}& 
{\bf 41.6, 25.8}$^{3,9}$\nl
$15030+4835$        & 64900 & 15.8 & 12.6$^B$& 3.5, 11.0$^9$    \nl
{\bf Mrk848,IZw107}& {\bf 12043} & {\bf 7.2} & {\bf 7.0$^A$} & 
{\bf 6.5,  4.8}$^{2,9}$  \nl
{\bf NGC6090,Mrk496} & {\bf 8754}  & {\bf 3.0} & {\bf 5.0$^C$} & 
{\bf 6.5,  3.5}$^2$      \nl
$16334+4630$        & 57270 & 16.6 & 9.6$^B$ & 4.4, 12.6$^9$    \nl
$16487+5447$        & 31250 & 15.2 & 14.5$^A$& 3.1, 5.3$^5$    \nl
$17132+5313$        & 15290 &  7.7 & 7.2$^A$ & 6.5,  6.4$^{1,3}$ \nl
$17208-0014$        & 12836 & 23.4 & 5.1$^B$ & 3.5,  2.7$^{7,9}$ \nl
{\bf NGC6670}       & {\bf  8684} & {\bf 3.8} & {\bf 11.6$^A$}&
{\bf 26.5, 14.6}$^9$  \nl
$19254-7245$/Sup-antena  & 18500 & 11.5& 7.7$^D$&9.3, 10.0$^7$\nl
$19297-0406$        & 25674 & 24.6 & 9.4$^B$ & 8.0, 11.5$^{7,9}$ \nl
20010$-$2352        & 15249 &  4.7 & 6.4$^A$ & 8.9,  8.1$^{10}$  \nl
{\bf 20550$+$1655/IIZw96}      & {\bf 10900} & {\bf 6.6} & {\bf 6.0$^A$} & 
{\bf 11.0, 7.4}$^{9,11}$   \nl
$20551-4250$   & 12900 &  9.6 & 4.8$^D$ & \nodata, 6.0$^7$\nl
$22491-1808$        & 23312 & 14.4 & 6.8$^C$ & 1.6, 2.4$^3$ \nl
$23128-5919$    & 13371 & 9.3 & 3.9$^D$ & 4.5, 4.0$^7$ \nl
{\bf NGC7592,Mrk928} & {\bf  7328} & {\bf 2.4} & {\bf 4.6$^A$} & 
{\bf 11.5, 5.3}$^{2,9}$    \nl
\tablenotetext{a}{A: our data; B: S97; 
C: Sanders \etal (1991); D: Mirabel \etal (1990); E: Kaz\'es \etal (1990);
F: Mirabel \etal (1989); G: Sanders \etal (1989). 
CO luminosity is recalculated using the consistent formula 
in S97, $L_l={\rm K~\kms~pc}^2$.}
\tablenotetext{b}{1: Armus \etal 1987; 2: Mazzarella \&
Boroson 1993; 3: Sanders 1992; 4: Kollatschny \etal 1991;
5: Murphy et al. 1996; 6: Klaas \& Els\"asser 1993;
7: Melnick \& Mirabel 1990; 8: Surace \etal 1998;
9: Fig.~1, S97 \& priv. com.; 10: van den Broek \etal 1991; 
11: Goldader \etal 1997.}
\tablecomments{Sources in the nearby complete sample have been
highlighted in boldface.}
\enddata
\end{deluxetable}

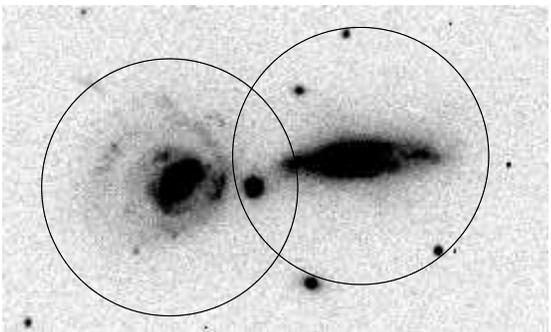
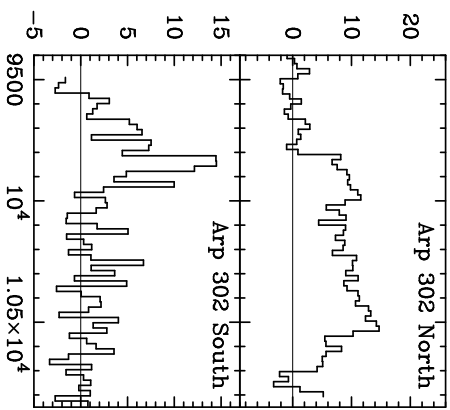
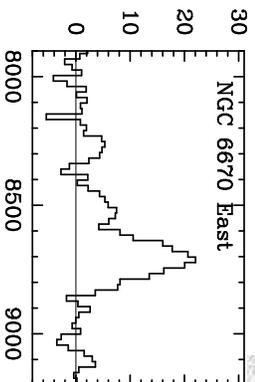
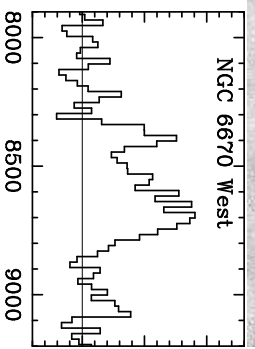
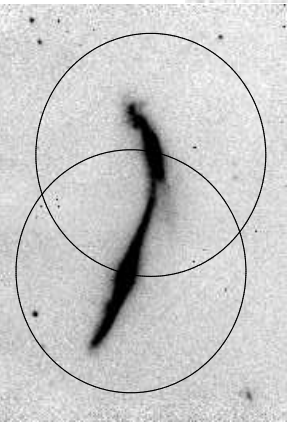
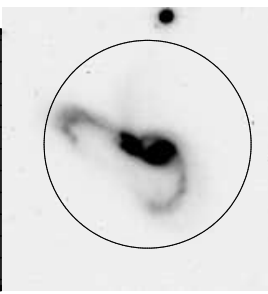
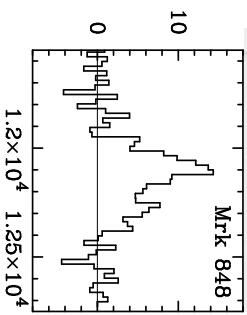
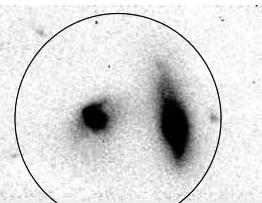
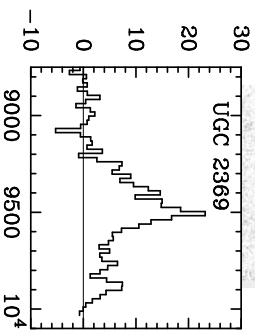
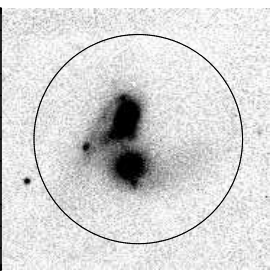
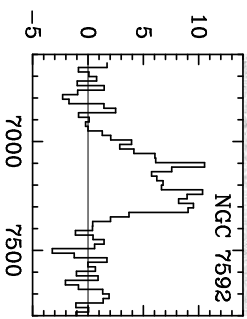
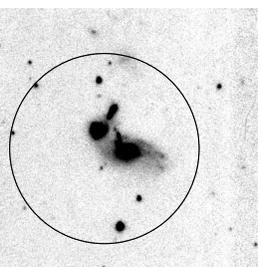
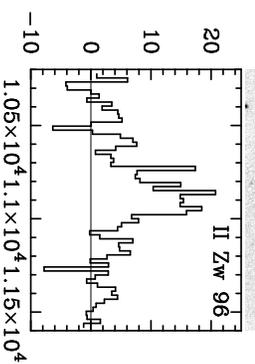
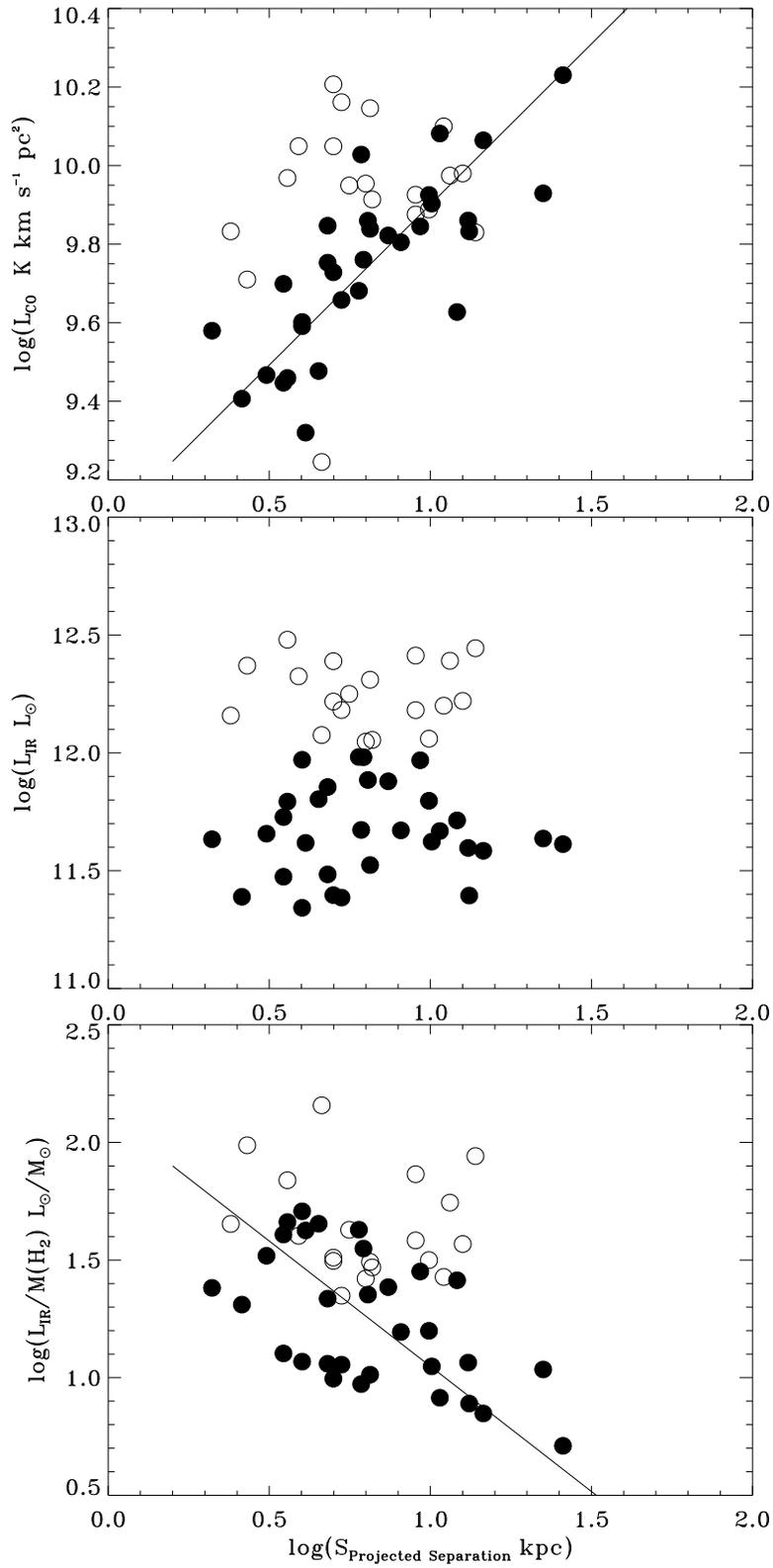
\end{document}